\documentstyle[preprint,aps,epsfig]{revtex}

\def\ep{\epsilon}
\def\tep{\tilde{\epsilon}}
\def\tbep{\tilde{\bar{\epsilon}}}
\def\bep{\bar{\epsilon}}
\def\bc{\bar{\chi}}

\begin{document}

\draft

\title{Nonlinear alternating current responses of graded materials}
\author{J. P. Huang$^1$, L. Gao$^{1,2}$, K. W. Yu$^1$, G. Q. Gu$^{1,3}$}

\address{$^1$Department of Physics, The Chinese University of Hong Kong,\\
 Shatin, New Territories, Hong Kong \\
$^2$Department of Physics, Suzhou University, Suzhou 215 006, China\\
$^3$College of Information Science and Technology,
 East China Normal University, \\ Shanghai 200 062, China }

\maketitle

\begin{abstract}
When a composite of nonlinear particles suspended in a host medium 
is subjected to a sinusoidal electric field, the electrical response 
in the composite will generally consist of alternating current (AC) 
fields at frequencies of higher-order harmonics. The situation becomes 
more interesting when the suspended particles are graded, with a spatial 
variation in the dielectric properties. 
The local electric field inside the graded particles can be calculated 
by the differential effective dipole approximation, which agrees 
very well with a first-principles approach. 
In this work, a nonlinear differential effective dipole approximation and 
a perturbation expansion method have been employed to investigate the 
effect of gradation on the nonlinear AC responses of these composites.
The results showed that the fundamental and third-harmonic AC responses 
are sensitive to the dielectric-constant and/or nonlinear-susceptibility 
gradation profiles within the particles. 
Thus, by measuring the AC responses of the graded composites, it is 
possible to perform a real-time monitoring of the fabrication process 
of the gradation profiles within the graded particles.
{\bf PACS number(s):} 77.84.Lf, 77.22.Ej, 42.65.-k, 42.79.Ry
\end{abstract}


\newpage
\section{Introduction}

Graded materials with spatial gradients in their
structure~\cite{Milton02} are abundant in nature, which have received
much attention as one of the advanced inhomogeneous composite
materials in diverse engineering applications~\cite{Meet90}. These
materials can be made to realize quite different physical properties 
from the homogeneous materials, and thus, to some extent, more useful 
and interesting. For graded materials, the traditional 
theories~\cite{Jackson75} for
homogeneous materials do not work any longer. Recently, we
presented a first-principles approach~\cite{Dong03,Gu03} and a
differential effective dipole approximation
(DEDA)~\cite{Yu-Un,Huang03}, to investigate the dielectric
properties of the graded materials. To our interest, the two
methods have been demonstrated in excellent agreement between each
other~\cite{Dong03}. In the case of graded materials, the problem
will become more complicated by the presence of nonlinearity
inside them. Fortunately, for deriving the equivalent nonlinear
susceptibility of graded particles, we have succeeded in putting
forth a nonlinear differential effective dipole approximation
(NDEDA)~\cite{Gao03-N}. As expected, this NDEDA has also been
demonstrated in excellent agreement with a first-principles
approach~\cite{Gao03-N}.

In addition, the finite-frequency response of nonlinear composite
materials have attracted much attention both in research and
industrial applications during the last two
decades~\cite{Bergman92}. When a composite with linear/nonlinear
particles embedded in a linear/nonlinear host medium is subjected
to a sinusoidal electric field, the electrical response in the
composite will generally consist of alternating current (AC)
fields at frequencies of higher-order
harmonics~\cite{Levy95,Hui98,Gu00,Huang01,HuangJAP03}. In fact, a
convenient method of probing the nonlinear characteristics of the
composite is to measure the harmonics of the nonlinear
polarization under the application of a sinusoidal electric
field~\cite{DJK98}. The strength of the nonlinear polarization
should be reflected in the magnitude of the harmonics. For the
purpose of extracting such harmonics, the perturbation
expansion~\cite{Gu00,Huang01,HuangJAP03} and self-consistent
methods~\cite{Huang01,Wan01} can be used.

In this work, based on the NDEDA, we shall investigate the effect
of gradation (inhomogeneity) inside the particles (inclusions) on
the AC responses of the graded composite by making use of a
perturbation expansion method~\cite{Gu92}. Here, the composite
under consideration is composed of linear/nonlinear graded
particles which are randomly embedded in a linear/nonlinear host
medium in the dilute limit. To this end, it is shown that the
fundamental and third-order harmonic AC responses are sensitive to
the dielectric-constant (or nonlinear-susceptibility) gradation
profile within the particle. Thus, by measuring the AC responses
of the graded composites, it is possible to perform a real-time
monitoring of the fabrication process of the gradation profiles
within graded particles.

This paper is organized as follows. In the following section, we
shall present the formalism, which is followed by the numerical
results in Section~\ref{NumericalResults}. In
Section~\ref{DisCon}, the discussion and conclusion will be given.

\section{Formalism}\label{Formalism}

Let's consider nonlinear graded spherical particles with radius
$a$ and dielectric gradation profile $\tilde{\ep}_1(r) =
\ep_1(r)+\chi_1(r)E_1{}^2$ inside it, being embedded in a
nonlinear host medium of dielectric constant
$\tilde{\ep}_2=\ep_2+\chi_2E_2{}^2$, in the presence of a uniform
external electric field $E_0$ along $z-$axis. Here $\ep_1(r)$ or
$\ep_2$ ($\chi_1(r)$ or $\chi_2$) denotes the corresponding linear
dielectric constant (nonlinear susceptibility), $E_1$ and $E_2$
stands for the local electric field inside the particles and the
host medium, respectively. Note both gradation profiles $\ep_1(r)$
and $\chi_1(r)$ are radial functions where $r<a.$ Throughout the
paper, we shall consider the case of weak nonlinearity only, that
is, $\chi_1(r)E_1{}^2\ll \ep_1(r)$ and $\chi_2E_2{}^2\ll \ep_2.$

\subsection{Comparison between a differential effective
dipole approximation and a first-principles
approach}\label{Comparison}

Recently, we put forth a DEDA (differential effective dipole
approximation)~\cite{Yu-Un,Huang03} for calculating the equivalent
dielectric constant $\bep_1(r)$~\cite{Gao03-N} of the spherical
graded particle. This DEDA receives the form
\begin{equation}
\frac{\mathrm{d}\bep_1(r)}{\mathrm{d}r}=\frac{[\ep_1(r)-\bep_1(r)]\cdot[\bep_1(r)+2\ep_1(r)]}
{r\ep_1(r)}.\label{DEDA}
\end{equation}
Note Eq.~(\ref{DEDA}) is just the Tartar formula, derived for
assemblages of spheres with varying radial and tangential
conductivity~\cite{Milton02}. So far, the equivalent $\bep_1(r=a)$
for the whole graded particle can be calculated, at least
numerically, by solving the differential equation
[Eq.~(\ref{DEDA})], as long as $\ep_1(r)$ (dielectric-constant
gradation profile) is given. Once $\bep_1(r=a)$ is determined, we
can readily take one step forward to obtain the volume average of
the linear local electric field inside the particles as
\begin{equation}
\langle{\bf E}_1{}^{\mathrm{(lin)}}\rangle =
\frac{3\ep_2}{\bep_1(r=a)+2\ep_2}{\bf E}_0.\label{LocalField}
\end{equation}
Hence, the DEDA [Eq.~(\ref{DEDA})] offers a convenient way to
obtain the local electric field [Eq.~(\ref{LocalField})]. It is
worth remarking that the DEDA [Eq.~(\ref{DEDA})] is valid for
arbitrary gradation profiles.

To show the correctness of Eq.~(\ref{LocalField}), we shall
alternatively present a first-principles approach for calculating
the local electric field inside the particle. For this purpose,
let's take the power-law gradation profile ($\ep(r)=A\cdot
(r/a)^n$) as a model. For this profile, the potential within the
graded particle can be given by solving the electrostatic
equation, $\nabla\cdot(\ep_1(r)\nabla\Phi) = 0$~\cite{Dong03},
\begin{equation}
\Phi_1(r) = -\eta_1E_0r^s\cos\theta,\,\, r<a,
\end{equation}
where the coefficient $\eta_1$ is determined by performing
appropriate boundary conditions,
$$
\eta_1 = \frac{3a^{1-s}\ep_2}{sA+2\ep_2},
$$
and $s=[\sqrt{9+2n+n^2}-(1+n)]/2.$ Based on the relation between
the linear local electric field and the potential (${\bf
E}_1{}^{\mathrm{(lin)}}(r)=-\nabla\Phi_1(r)$), we have
\begin{eqnarray}
{\bf E}_1{}^{\mathrm{(lin)}}(r) &=&
\eta_1E_0r^{s-1}\{(s-1)\cos\theta\sin\theta\cos\phi\hat{{\bf
x}}+(s-1)\cos\theta\sin\theta\sin\phi\hat{{\bf y}} \nonumber\\
& &+[(s-1)\cos^2\theta+1]\hat{{\bf z}}\},\label{FP}
\end{eqnarray}
where $\hat{{\bf x}}, \hat{{\bf y}},$ and $\hat{{\bf z}}$ are the
unit vectors along $x-$, $y-$, and $z-$axes, respectively. So far,
it is straightforward to obtain the volume average of the local
electric field inside the particles,
\begin{equation}
\langle{\bf E}_1{}^{\mathrm{(lin)}}\rangle = \frac{1}{V}\int_V{\bf
E}_1{}^{\mathrm{(lin)}}(r)\mathrm{d}V,\label{Volume}
\end{equation}
where $V$ is the volume of the spherical particles.

In Fig.~1, we shall numerically compare Eq.~(\ref{LocalField})
(local field predicted by the DEDA) with Eq.~(\ref{Volume}) (local
field obtained from the first-principles approach).

\subsection{Nonlinear polarization and its higher harmonics}

\subsubsection{Nonlinear differential effective dipole
approximation}

In a recent work~\cite{Gao03-N}, we have established an NDEDA
(nonlinear differential effective dipole approximation), by
deriving a differential equation for the equivalent nonlinear
susceptibility $\bar{\chi}_1(r)$, namely,
\begin{eqnarray}
\frac{{\rm d}\bar{\chi}_1(r)}{{\rm
d}r}&=&\bar{\chi}_1(r)\left[\frac{4{\rm d}\bar{\epsilon}_1(r)/{\rm
d}r}{2\epsilon_2+ \bar{\epsilon}_1(r)} \right]+
\bar{\chi}_1(r)\cdot\frac{8y-3}{r} +\frac{3\chi_1(r)}{5r}
\nonumber \\
&&\cdot
\left(\frac{\bar{\epsilon}_1(r)+2\epsilon_1(r)}{3\epsilon_1(r)}\right)^4(5+36x^2+16x^3+24x^4).\label{NDEDA1}
\end{eqnarray}
where
$$
x=\frac{\bep_1(r)-\ep_1(r)}{\bep_1(r)+2\ep_1(r)}\,\,\mathrm{and}\,\,
y=\frac{[\ep_1(r)-\ep_2]\cdot[\bep_1(r)-\ep_1(r)]}{\ep_1(r)[\bep_1(r)+2\ep_2]}.
$$
Similarly, $\bar{\chi}_1(r=a)$ can be obtained, at least
numerically, by solving Eq.~(\ref{NDEDA1}), once the initial
conditions, namely, $\ep_1(r=0)$ and $\chi_1(r=0)$ are given.

In what follows, we can investigate the nonlinear AC response of
the graded spherical particle by seeing it as a homogeneous
particle having the constitutive relation between the displacement
and the local electric field,~\cite{Stroud88}
$$
{\bf D}_1 = \bep_1(r=a){\bf E}_1+\bar{\chi}_1(r=a)E_1{}^2{\bf
E}_1\equiv \tbep_1(r=a){\bf E}_1,
$$
where $\bep_1(r=a)$ and $\bar{\chi}_1(r=a)$ are determined by
Eqs.~(\ref{DEDA})~and~(\ref{NDEDA1}), respectively. For the sake
of convenience, we shall represent $\bep_1(r=a)$ by $\bep_1$,
$\tbep_1(r=a)$ by $\tbep_1$ as well as $\bc_1(r=a)$ by $\bc_1$, if
no special instructions.

\subsubsection{Nonlinear AC responses}

If we apply a sinusoidal electric field like
\begin{equation}
E_0(t)=E_0\sin(\omega t),\label{Time-depend}
\end{equation}
the local electric field $\sqrt{\langle E_1{}^2\rangle}$ and the
induced dipole moment
\begin{equation}
\tilde{p}=\tilde{\ep}_ea^3\frac{\tbep_1-\tilde{\ep}_2}{\tbep_1+2\tilde{\ep}_2}E_0\label{DM}
\end{equation}
will depend on time sinusoidally, too. Here the effective
dielectric constant of the system ($\tep_e$) is given by the
following dilute-limit expression,
\begin{equation}
\tep_e=\tep_2+3\tep_2f\frac{\tbep_1-\tep_2}{\tbep_1+2\tep_2},\label{DL-Expre}
\end{equation}
where $f$ is the volume fraction of the particles. By virtue of
the inversion symmetry, the local electric field is a
superposition of odd-order harmonics such that
\begin{equation}
\sqrt{\langle E_1{}^2\rangle} = E_{\omega}\sin(\omega
t)+E_{3\omega}\sin(3\omega t)+\cdots.
\end{equation}
Similarly, the induced dipole moment contains harmonics as
\begin{equation}
\tilde{p} = p_{\omega}\sin(\omega t)+p_{3\omega}\sin(3\omega
t)+\cdots .
\end{equation}
These harmonics coefficients can be extracted from the time
dependence of the solution of $\sqrt{\langle E_1{}^2\rangle}$ and
$\tilde{p}.$

\subsubsection{Analytic solutions for the nonlinear AC responses}

In what follows, we will perform a perturbation expansion method
to extract the third harmonics of the local electric field and the
induced dipole moment. It is known that the perturbation expansion
method is applicable to weak nonlinearity only, limited by the
convergence of the series expansion.

Let's start from the dilute-limit expression for the effective
linear dielectric constant ($\ep_e$) of the system of interest,
namely, Eq.~(\ref{DL-Expre}) where $\bc_1=\chi_2=0$.

Next, we obtain the linear local electric fields $\langle
E_1{}^2\rangle$ and $\langle E_2{}^2\rangle$, respectively,
\begin{eqnarray}
\langle E_1{}^2\rangle &=&
\frac{E_0{}^2}{f}\frac{\partial\ep_e}{\partial\bep_1}\equiv
F(\bep_1,\ep_2,f,E_0),\\
\langle E_2{}^2\rangle &=&
\frac{E_0{}^2}{1-f}\frac{\partial\ep_e}{\partial\ep_2}\equiv
G(\bep_1,\ep_2,f,E_0).
\end{eqnarray}
In view of the existence of nonlinearity inside the two
components, we readily obtain the following local electric fields
for the nonlinear particle and host, respectively,
\begin{eqnarray}
\langle E_1{}^2\rangle &=&
F(\tbep_1,\tilde{\ep}_2,f,E_0),\label{NonlinearE1}\\
\langle E_2{}^2\rangle &=&
G(\tbep_1,\tilde{\ep}_2,f,E_0).\label{NonlinearE2}
\end{eqnarray}

For the below series expansions, we will take
$\tbep_1=\bep_1+\bc_1E_1{}^2\approx \bep_1+\bc_1\langle
E_1{}^2\rangle$ and $\tep_2=\ep_2+\chi_2E_2{}^2\approx
\ep_2+\chi_2\langle E_2{}^2\rangle,$ in
Eqs.~(\ref{NonlinearE1})~and~(\ref{NonlinearE2}), where
$\langle\cdots\rangle$ denotes the volume average of $\cdots.$
Let's expand the local electric field $\langle E_1{}^2\rangle$ and
$\langle E_2{}^2\rangle$ into a Taylor expansion, taking
$\bc_1\langle E_1{}^2\rangle$ and $\chi_2\langle E_2{}^2\rangle$
as the perturbative quantities,
\begin{eqnarray}
\langle E_1{}^2\rangle &=&
F(\bep_1,\ep_2,f,E_0)+\frac{\partial}{\partial\tbep_1}F(\tbep_1,\ep_2,f,E_0)|_{\tbep_1=\bep_1}\bc_1\langle E_1{}^2\rangle+\nonumber\\
& &
\frac{\partial}{\partial\tep_2}F(\bep_1,\tep_2,f,E_0)|_{\tep_2=\ep_2}\chi_2\langle E_2{}^2\rangle+\cdots,\label{expan-E1}\\
\langle E_2{}^2\rangle &=&
G(\bep_1,\ep_2,f,E_0)+\frac{\partial}{\partial\tbep_1}G(\tbep_1,\ep_2,f,E_0)|_{\tbep_1=\bep_1}\bc_1\langle E_1{}^2\rangle+\nonumber\\
& &
\frac{\partial}{\partial\tep_2}G(\bep_1,\tep_2,f,E_0)|_{\tep_2=\ep_2}\chi_2\langle
E_2{}^2\rangle+\cdots.
\end{eqnarray}
Keeping the lowest orders of $\bc_1\langle E_1{}^2\rangle$, we can
rewrite Eq.~(\ref{expan-E1}) as,
\begin{equation}
\langle E_1{}^2\rangle = h_1E_0{}^2+(h_2+h_3)E_0{}^4,
\end{equation}
where
\begin{eqnarray}
h_1 &=& \frac{9\ep_2{}^2}{(\bep_1+2\ep_2)^2},\,\,
h_2 = -\frac{162\ep_2{}^4\bc_1}{(\bep_1+2\ep_2)^5},\nonumber\\
h_3 &=&
\frac{18\bep_1\ep_2\chi_2[(1+3f)\bep_1{}^2+(4-6f)\bep_1\ep_2+(4-6f)\ep_2{}^2]}{(1-p)(\bep_1+2\ep_2)^5}.\nonumber
\end{eqnarray}

Because of the time-dependence of the electric field
[Eq.~(\ref{Time-depend})], we can take one step forward to obtain
 the local electric field in terms of the harmonics ($E_{\omega}$
and $E_{3\omega}$),
\begin{equation}
\sqrt{\langle E_1{}^2\rangle} = E_{\omega}\sin(\omega
t)+E_{3\omega}\sin(3\omega t),
\end{equation}
where
\begin{eqnarray}
E_{\omega} &=&
\sqrt{h_1}E_0+\frac{3}{8}\frac{h_2+h_3}{\sqrt{h_1}}E_0{}^3,\\
E_{3\omega} &=& -\frac{1}{8}\frac{h_2+h_3}{\sqrt{h_1}}E_0{}^3.
\end{eqnarray}

Similarly, based on Eq.~(\ref{DM}), we obtain the induced dipole
moment in terms of the harmonics ($p_{\omega}$ and $p_{3\omega}$),
\begin{equation}
\tilde{p}/a^3 = (p_{\omega}/a^3)\sin(\omega
t)+(p_{3\omega}/a^3)\sin(3\omega t),
\end{equation}
where
\begin{eqnarray}
p_{\omega}/a^3 &=&
k_1E_0+\frac{3}{4}(k_2+k_3)E_0{}^3,\\
p_{3\omega}/a^3 &=& -\frac{1}{4}(k_2+k_3)E_0{}^3,
\end{eqnarray}
with

\begin{eqnarray}
k_1 &=& \ep_e\frac{\bep_1-\ep_2}{\bep_1+2\ep_2},\,\,
k_2 = \frac{3\ep_2{}^2h_1\bar{\chi}_1[\bep_1+6f\bep_1+(2-6f)\ep_2]}{(\bep_1+2\ep_2)^3},\nonumber\\
k_3 &=&
\frac{j_1\chi_2[(1+3f)\bep_1{}^3-18f\bep_1{}^2\ep_2-3(2-3f)\bep_1\ep_2{}^2-2(2-3f)\ep_2{}^3]}
{(\bep_1+2\ep_2)^3}\nonumber.
\end{eqnarray}

%
In
the above derivation, we have used an identity $ \sin^3(\omega
t)=(3/4)\sin(\omega t)-(1/4)\sin(3\omega t).$

\section{Numerical results}\label{NumericalResults}

For numerical calculations, we take $\chi_1(r)=\chi_1(0)+D(r/a)$,
and other parameters: volume fraction $p=0.09,$ external field
strength $E_0=1,$ linear part of host dielectric constant
$\ep_2=1.$

The validation of using the DEDA is shown in Fig.~1. In this
figure, we investigate the linear local electric field by using a
power-law gradation profile inside the particles, in an attempt to
compare the DEDA with the first-principles approach. As expected,
the excellent agreement is demonstrated between the DEDA
[Eq.~(\ref{LocalField})] and the first-principles approach
[Eq.~(\ref{Volume})]. In addition, it is worth noting that, for a
linear gradation profile within the graded particles, the
first-principles approach holds as well~\cite{Dong03}, and the
excellent agreement between the two methods can also be found
(figure not shown here).

Next, we discuss a power-law gradation profile $[\ep_1(r)=A\cdot
(r/a)^n],$ see Fig.~2. In this figure, the harmonics of local
electric field and the induced dipole moment are investigated as a
function of $A$ for various $n.$ In this case, increasing $A$ (or
decreasing $n$) leads to increasing $\tbep_1$ (namely, the
equivalent dielectric constant of the graded particle under
consideration) and in turn yields a decreasing local electric
field inside the particle. Thus, either an increase in $A$ or a
decrease in $n$ leads to the weakening third-order harmonics
$(E_{3\omega}$ and $p_{3\omega}),$ as displayed in Fig.~2.

The $x-$axes of Figs.~3~and~4 represent the slope ($C$) of a
linear gradation profile. It is because during the fabrication of
graded spherical particles by using diffusion, the dielectric
constant at the center $\ep_1(0)$ may vary slightly while that at
the grain boundary can change substantially.

For a linear gradation profile $[\ep_1(r)=\ep_1(0)+C\cdot (r/a)],$
Fig.~3 shows the harmonics as a function of $C$ for various
$\ep_1(0).$ In this case, increasing $C$ or $\ep_1(0)$ yield
increasing $\tbep_1,$ and hence one obtains the decreasing local
electric field. As a result, larger $C$ or $\ep_1(0),$ weaker the
third-order harmonics $(E_{3\omega}$ and $p_{3\omega}),$ see
Fig.~3.

Fig.~4 displays the effect of $\chi_2$ on the harmonics, for a
linear gradation profile $[\ep_1(r)=\ep_1(0)+C\cdot (r/a)].$ Here,
increasing $\chi_2$ leads to increasing the local electric field
inside the graded particle of interest. Therefore, the third-order
harmonics $(E_{3\omega}$ and $p_{3\omega})$ increase for
increasing $\chi_2.$

As mentioned above, as $A$ and $C$ increases, the equivalent
dielectric constant of the particle should be increased
accordingly, which in turn yields a decreasing local electric
field, and hence, in Figs.~2$\sim$4, $E_{\omega}$ decreases for
increasing $A$ or $C.$ On the other hand, it is found that, in
Figs.~2$\sim$4, $p_{\omega}$ increases for increasing $A$ or $C$
which is, in fact, due to the increasing effective dielectric
constant $\tep_e$ [refer to Eq.~(\ref{DM})]. Similarly, this
analysis works fairly for understanding the effect of $n$ and
$\ep_1(0)$ on $E_{\omega}$ and $p_{\omega}$, as displayed in
Figs.~2~and~3. However, increasing $\chi_2$ can increase not only
the local electric field inside the particles, but also the
effective dielectric constant $\tep_e,$ and hence we observe
increasing $E_{\omega}$ and $p_{\omega},$ as shown in Fig.~4.

In addition, we also discuss the effect of
nonlinear-susceptibility gradation profiles (no figures shown
here). For linear gradation profile $\chi_1(r)=\chi_1(0)+D(r/a)$,
as $\chi_1(0)$ (or $D$) increases, the third harmonics of both the
electric field and the induced dipole moment increases
accordingly. On the other hand, for pow-law gradation profile
$\chi_1(r)=B(r/a)^m$, increasing $B$ (or decreasing $m$) leads to
increasing third harmonics. To understand such results, we can
again resort to the above analysis on the effect of the relevant
parameters on the local field as well as the effective dielectric
constant.

\section{Discussion and conclusion}\label{DisCon}

Here some comments are in order. We investigate the nonlinear AC
responses of the graded material where linear/nonlinear graded
particles are randomly embedded in a linear/nonlinear host medium
in the dilute limit. As a matter of fact, the NDEDA (nonlinear
differential effective dipole approximation) is valid for
arbitrary gradation profiles, besides the power-law and linear
profiles of interest. In particular, based on the first-principles
approach, the exact solution is obtainable, for not only power-law
profiles (see Section~\ref{Comparison}), but also linear profiles
(refer to Ref.~\cite{Dong03}).

As an extension, it is of particular interest to see what happens
to the nonlinear AC responses of graded particles in
electrorheological fluids, in which a field-induced anisotropic
structure often occurs. For discussing this anisotropy effect, we
can make use of the Maxwell-Garnett approximation for anisotropic
structures~\cite{Huang01}.

To sum up, based on our recently-established NDEDA, we have
investigated the nonlinear AC responses of a composite with
linear/nonlinear graded spherical particles embedded in a
linear/nonlinear host medium, and found the fundamental and third
harmonic AC responses are sensitive to the dielectric-constant (or
nonlinear-susceptibility) gradation profile within the particles.
Again, for extracting the linear local electric field, the DEDA
agrees very well with the first-principles approach. Thus, by
measuring the AC responses of the graded composites, it is
possible to perform a real-time monitoring of the fabrication
process of the gradation profiles within graded particles.

\section*{Acknowledgments}

This work has been supported by the Research Grants Council of the
Hong Kong SAR Government under project numbers CUHK 4245/01P and
CUHK 403303, the National Natural Science Foundation of China
under Grant No.~10204017 (L.G.), the  Natural Science of Jiangsu
Province under Grant No.~BK2002038 (L.G.), and National ``863''
Project of China under grant number 2002AA639270 (G.Q.G.).

\newpage

\begin{figure}[ht]
\caption{For power-law gradation profile $\ep_1(r)=A\cdot
(r/a)^n,$ comparison between the approximation result [obtained
from the DEDA, Eq.~(\ref{LocalField})] and the exact solution
[predicted by a first-principles approach, Eq.~(\ref{Volume})],
for linear electric field $E_1{}^{\rm{(lin)}}$ as a function of
$A$ for various $n$. Parameters: $\chi_1(0)=0.1$, $D=0.1,$
$\chi_2=0.$}
\end{figure}

\begin{figure}[h]
\caption{For power-law gradation profile $\ep_1(r)=A\cdot
(r/a)^n,$ harmonics of the local electric field and induced dipole
moment, as a function of $A$, for various $n$.  Parameters:
$\chi_1(0)=0.1,$ $D=0.1,$ $\chi_2=0.$}
\end{figure}

\begin{figure}[h]
\caption{For linear gradation profile $\ep_1(r)=\ep_1(0)+C\cdot
(r/a),$ harmonics of the local electric field and induced dipole
moment, as a function of $C$, for various $\ep_1(0)$.  Parameters:
$\chi_1(0)=0.1,$ $D=0.1,$ $\chi_2=0.$}
\end{figure}

\begin{figure}[h]
\caption{Same as Fig.~3, but for various $\chi_2$. Parameters:
$\chi_1(0)=0,$ $D=0,$ $\ep_1(0)=3.$}
\end{figure}

\newpage
\centerline{\epsfig{file=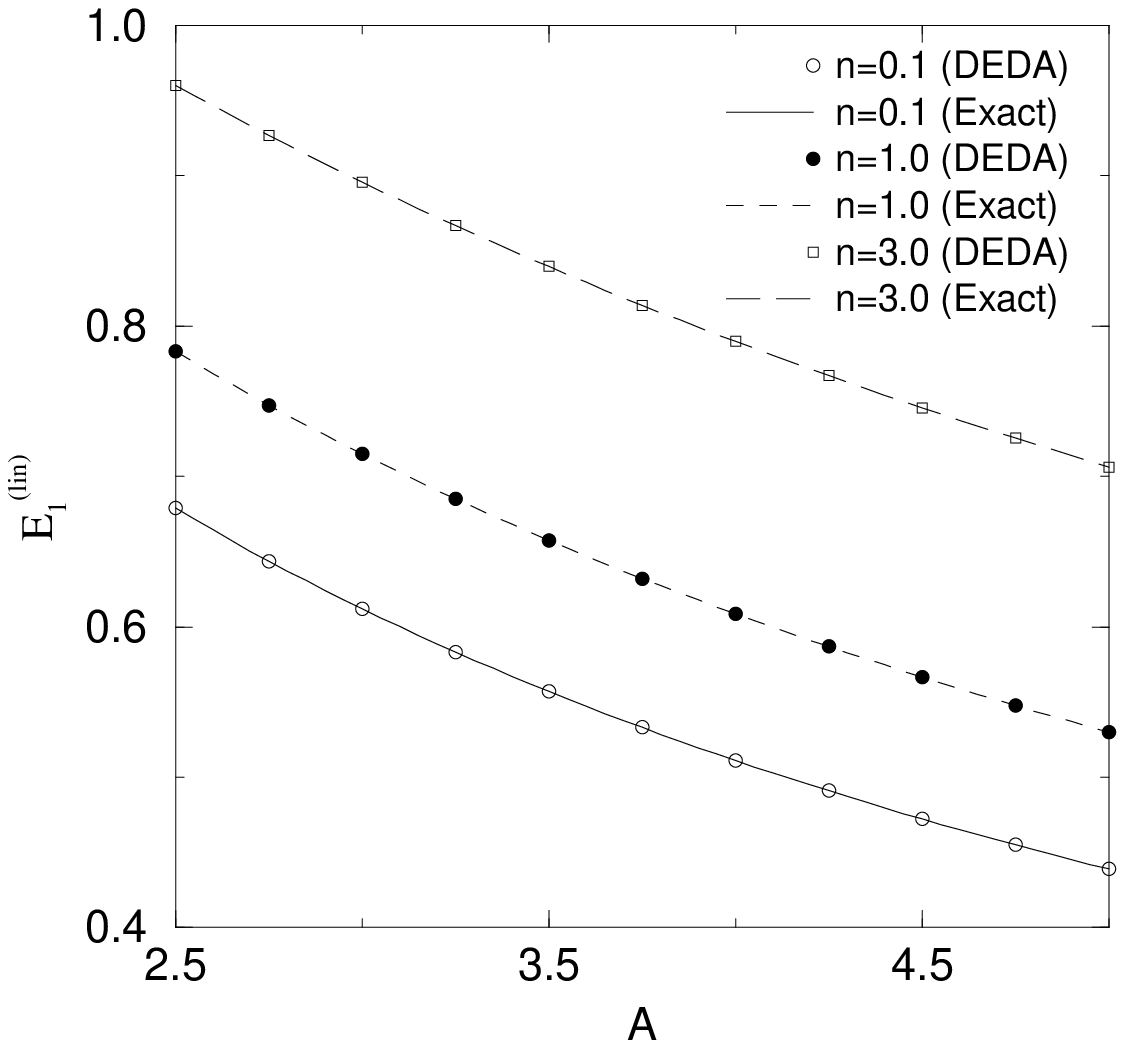,width=250pt}}
\centerline{Fig.1}

\newpage
\centerline{\epsfig{file=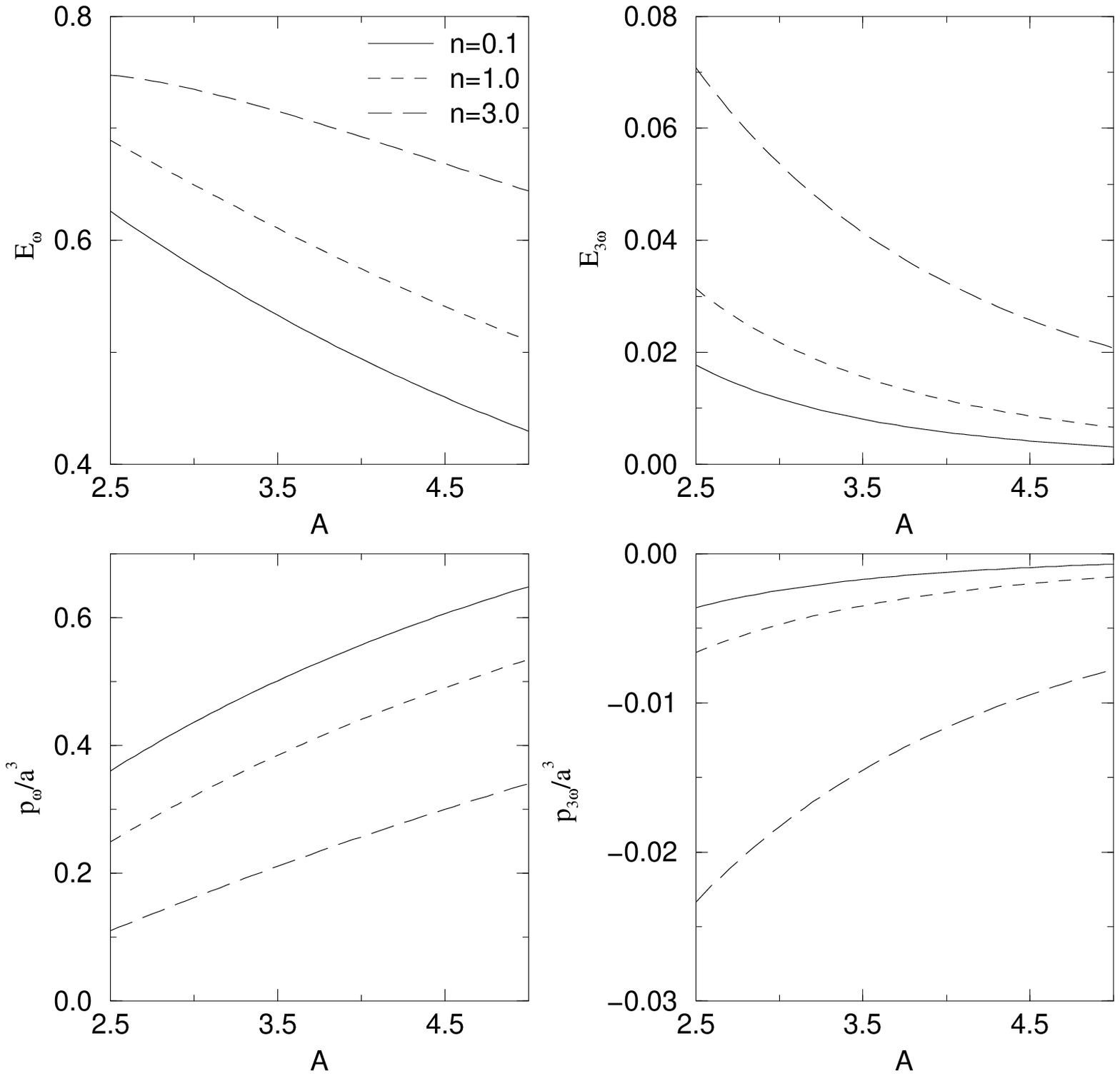,width=250pt}}
\centerline{Fig.2}

\newpage
\centerline{\epsfig{file=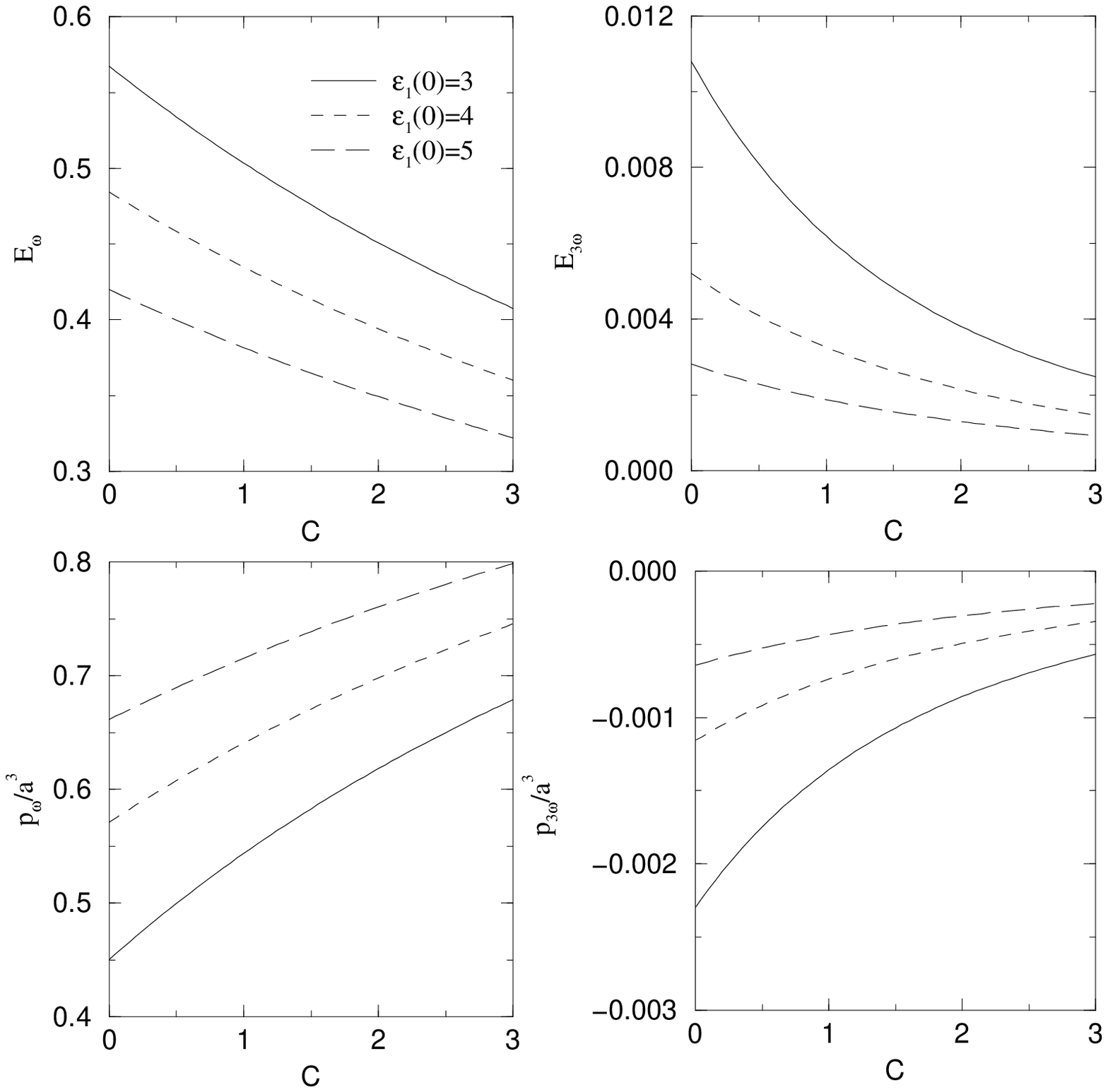,width=250pt}}
\centerline{Fig.3}

\newpage
\centerline{\epsfig{file=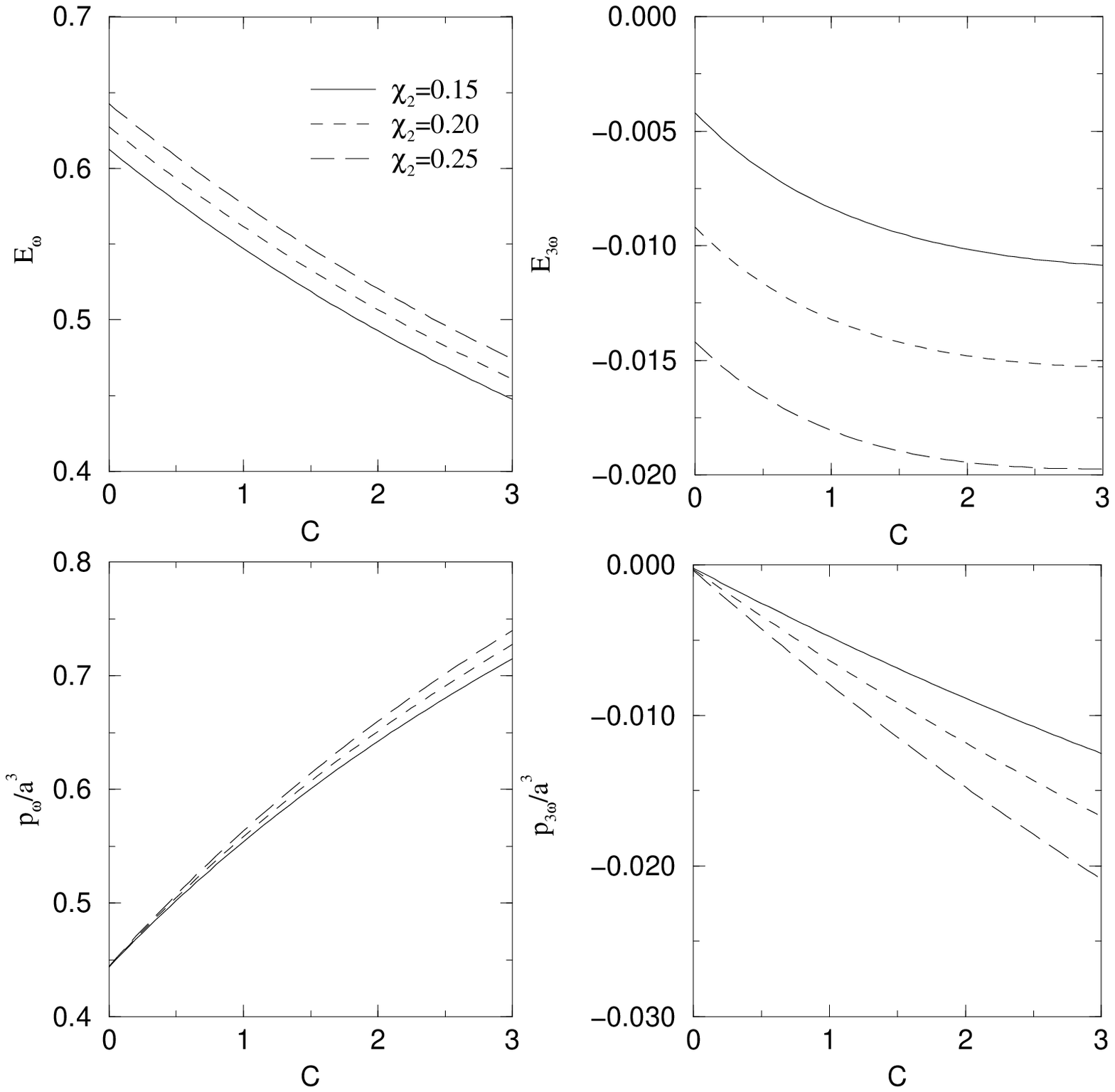,width=250pt}}
\centerline{Fig.4}


\begin{references}

\bibitem{Milton02} G. W. Milton, {\it The Theory of Composites},
Chapter 7 (Cambridge University Press, Cambridge, 2002).


\bibitem{Meet90} M. Yamanouchi, M. Koizumi, T. Hirai and I.
Shioda, in {\it Proceedings of the First International Symposium
on Functionally Graded Materials} (Sendi, Japan, 1990).

\bibitem{Jackson75} J. D. Jackson, {\it Classical Electrodynamics}
(Wiley, New York, 1975).

\bibitem{Dong03} L. Dong, G. Q. Gu, and K. W. Yu, Phys. Rev. B {\bf
67}, 224205 (2003).

\bibitem{Gu03} G. Q. Gu and K. W. Yu, J. Appl. Phys., to be
published on Sep. 15, 2003.

\bibitem{Yu-Un} K. W. Yu, G. Q. Gu and J. P. Huang, preprint: cond-mat/0211532.

\bibitem{Huang03} J. P. Huang, K. W. Yu, G. Q. Gu, and M.
Karttunen, Phys. Rev. E {\bf 67}, 051405 (2003).

\bibitem{Gao03-N} L. Gao, J. P. Huang, and K. W. Yu, preprint: cond-mat/0308032.


\bibitem{Bergman92} D. J. Bergman and D. Stroud, Solid State Phys. {\bf 46},
147 (1992).

\bibitem{Levy95} O. Levy, D. J. Bergman, and D. Stroud, Phys. Rev. E {\bf 52},
 3184 (1995).

 \bibitem{Hui98} P. M. Hui, P. C. Cheung, and D. Stroud, J. Appl. Phys. {\bf 84}, 3451 (1998).

\bibitem{Gu00} G. Q. Gu, P. M. Hui, and K. W. Yu, Physica B {\bf
279}, 62 (2000).

 \bibitem{Huang01} J. P. Huang, J. T. K. Wan, C. K. Lo, and K. W. Yu,
Phys. Rev. E {\bf 64}, 061505(R) (2001).

 \bibitem{HuangJAP03} J. P. Huang, L. Gao, and K. W. Yu, J. Appl. Phys. {\bf 93}, 2871 (2003).

\bibitem{DJK98} D. J. Klingenberg, MRS Bull. {\bf 23}, 30 (1998).

\bibitem{Wan01} J. T. K. Wan, G. Q. Gu, and K. W. Yu, Phys. Rev. E
{\bf 63}, 052501 (2001).

\bibitem{Gu92} G. Q. Gu and K. W. Yu, Phys. Rev. B {\bf 46}, 4502 (1992).


\bibitem{Stroud88} D. Stroud and P. M. Hui, Phys. Rev. B {\bf 37}, 8719 (1988).







\end{references}
\end{document}